\documentclass{aastex63}

\received{\today}
\revised{\today}
\accepted{\today}
\submitjournal{ApJL}

\shorttitle{GD\,394 and TESS}
\shortauthors{Wilson et al. 2020}
\graphicspath{{./}{figures/}}

\begin{document}


\title{Optical detection of the 1.1-day variability at the white dwarf GD\,394 with \textit{TESS}}

\correspondingauthor{David J. Wilson}
\email{djwilson394\@gmail.com}

\author[0000-0001-9667-9449]{David J. Wilson}
\affil{McDonald Observatory, University of Texas at Austin, Austin, TX 78712}

\author{J. J. Hermes}
\affil{Department of Astronomy, Boston University, 725 Commonwealth Ave., Boston, MA 02215, USA}

\author{Boris T. G{\"a}nsicke}
\affil{Department of Physics, University of Warwick, Coventry CV4 7AL,UK}



\begin{abstract}
Recent discoveries have demonstrated that planetary systems routinely survive the post-main sequence evolution of their host stars, leaving the resulting white dwarf with a rich circumsteller environment. Among the most intriguing of such hosts is the hot white dwarf GD\,394, exhibiting a unique $1.150\pm0.003$\,d flux variation detected in \textit{Extreme Ultraviolet Explorer} (\textit{EUVE}) observations in the mid 1990s. The variation has eluded a satisfactory explanation, but hypotheses include channeled accretion producing a dark spot of metals, occultation by a gas cloud from an evaporating planet, or heating from a flux tube produced by an orbiting iron-cored planetesimal. 

We present observations obtained with the \textit{Transiting Exoplanet Survey Satellite} (\textit{TESS}) of GD\,394. The space-based optical photometry demonstrates a $0.12\pm0.01$\% flux variation with a period of $1.146\pm 0.001$\,d, consistent with the \textit{EUVE} period and the first re-detection of the flux variation outside of the extreme ultraviolet. We describe the analysis of the \textit{TESS} light curve and measurement of the optical variation, and discuss the implications of our results for the various physical explanations put forward for the variability of GD\,394.    

\end{abstract}



\section{Introduction} \label{sec:intro}
GD\,394 is a hot, metal-polluted white dwarf that has challenged interpretation since its initial identification by \citet{giclasetal67-1}. Indeed, both of the descriptions in the preceding sentence are unquantified: Estimates for the effective temperature vary from 33000--41000\,K \citep{lajoie+bergeron07-1, barstowetal96-1} and the measured metal abundances, accretion rates, and species depend on the wavelength band and ionisation levels observed \citep{wilsonetal19-1}. However, the most intriguing aspect of GD\,394 is the detection by \citet{christianetal99-1} and \citet{dupuisetal00-1} of a sustained 25\% modulation of the extreme ultraviolet (EUV 70--380\,\AA) flux with a period of 1.15\,d in observations made in 1992--1996 with multiple instruments onboard the {\em Extreme Ultraviolet Explorer} ({\em EUVE}) satellite. This large-amplitude EUV modulation is so far unique among white dwarfs, and was hypothesised to be due to opacity changes induced by a spot of accreting metals moving in to and out of view with the white dwarf rotation. 

The metal spot hypothesis made two observable predictions. First, the strength of the metal lines in the white dwarf spectrum should vary in phase with the EUV variation. Second, there should be an anti-phase flux variation at optical wavelengths due to flux redistribution. Follow-up observations  by \citet{wilsonetal19-1} ruled out the first of these predictions, finding no change in the strength of strong Si, Fe and Al absorption lines in eight \textit{Hubble Space Telescope} (\textit{HST}) far ultraviolet (FUV, 1160-1700\,\AA) spectra sampling the full (putative) white dwarf spin cycle. They also searched SuperWASP photometry for optical variability, ruling out $\gtrsim$\,1\% changes in flux. Instead of a spot model, \citet{wilsonetal19-1} favoured a circumstellar explanation such as a gas cloud generated by an evaporating but non-transiting planet, similar to the ice giant detected in an $\approx$8--10\,d orbit of WD J0914+1914 by \citet{gaensickeetal19-1}. \citet{veras+wolszczan19-1} alternatively suggested that an orbiting, iron-rich planetesimal core could induce a magnetic flux tube connecting it to the white dwarf in GD\,394, heating the photosphere at the base of the tube to produce a hot spot. 
 
Here, we present observations of GD\,394 obtained using the {\em Transiting Exoplanet Survey Satellite} \citep[\textit{TESS},][]{rickeretal14-1} demonstrating that GD\,394 is indeed varying in the optical with a period in agreement with that detected previously in the EUV, but with a much smaller amplitude. Section \ref{sec:obs} details our analysis of the \textit{TESS} light curve. Section \ref{sec:disc} discusses the implications of the optical re-detection for the physical explanation of the variation at GD\,394. 

\begin{figure}
    \centering
    \includegraphics[width=\columnwidth]{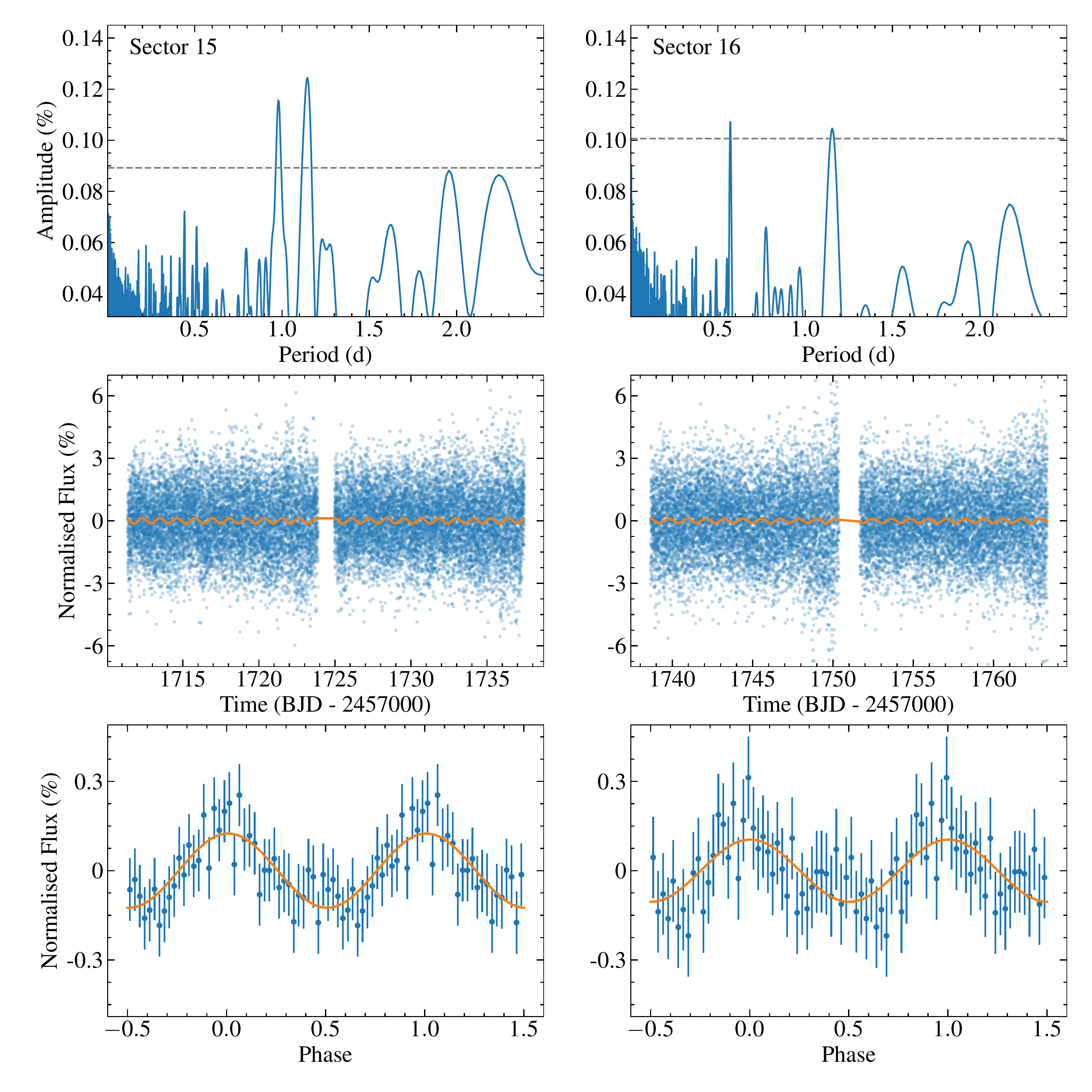}
    \caption{Top: Periodograms of the \textit{TESS} light curves of GD\,394 from Sectors 15 and 16. The grey dashed line shows the 99\% false alarm probability signal apparent in each sector, as discussed in the text. Significant signals at $1.1459\pm0.0033$\,d and $1.1547\pm0.0051$\,d  are detected in Sector 15 and 16 respectively. The $\approx 0.98$\,d signal apparent in Sector 15 is due to contamination from a nearby giant star. The first harmonic ($P$/2) of the 1.15\,d signal is detected in Sector 16. Middle: \textit{TESS} light curves of GD\,394 together with the sine fit used to measure the period and amplitude of the variation. The enhanced scatter at the end of each segment of the light curve is due to increased background Earthshine as the spacecraft approaches perigee. Bottom: Light curves folded onto the fitted period and binned to 40 steps. The cycle is repeated for clarity, and the model fit is overplotted in orange. }
    \label{fig:all_obs}
\end{figure}




\section{Observations} \label{sec:obs}


GD\,394 (TIC 259773610, $T$=13.4\,mag) was observed by {\em TESS} in Camera~2 for 52 days in Sectors 15 and 16 (2019~August~15--2019~October~06), with three roughly one-day gaps at spacecraft perigee\footnote{See \textit{TESS} Data Release Notes at \url{http://archive.stsci.edu/tess/tess_drn.html}}. Data was returned with a two~minute cadence as requested in proposals G022077, G022028 and G022017, and processed using the Pre-Search  Data  Conditioning  Pipeline  \citep[PDC,][]{stumpeetal12-1} to remove common known instrumental trends.

\defcitealias{lightkurve18-1}{Lightkurve Collaboration, 2018}

We analysed the light curves from each sector separately. The light curves were retrieved from MAST\footnote{https://archive.stsci.edu/} and points marked with a quality flag were removed, as were any 5-sigma outliers above the median flux. The flux was normalised by subtracting a second-order polynomial fit. We generated Lomb-Scargle periodograms using the \texttt{Lightkurve} package \citepalias{lightkurve18-1} (Figure \ref{fig:all_obs}, top). A $\approx 1.15$\,d period is clearly detected in each sector, providing the first confirmation of the {\em EUVE} detection beyond the extreme ultraviolet. 

The Sector 15 periodogram contained a second peak at $\approx 0.98$\,d (orange dashed line in Figure \ref{fig:all_obs}). Using the \textit{TESS} pixel data, we found that this signal originates from a nearby giant star, TIC 259773551 (\textit{TESS} magnitude = 12.9). We have therefore ignored this periodicity; it only marginally adds to our  systematic uncertainties.

To ascertain the significance of the detected signals we calculated the false alarm probability (FAP) via the method described in \citet{hermesetal17-1,belletal19-1}. In short, we generated $10{,}000$ synthetic light curves for each sector, using the same time-axis but randomly shuffling the flux values. A periodogram was calculated for each synthetic light curve and the maximum amplitude recorded. We defined our 1\% FAP as the power below which the maximum amplitude in 99\% of our synthetic light curves fell (grey dashed line in Figure \ref{fig:all_obs}). As the signals in each light curve exceed this limit, we conclude that there is a less than one per\,cent probability that our detected signal is due to random chance in each sector, and therefore a $<$0.01\% chance of observing the same, random signal in both sectors. A $P/2$ harmonic of the expected signal is also detected with $< 1\%$ FAP in Sector 16.

We then fit a sinusoid to each light curve using the peak measured from the periodogram as an initial guess for the period, shown in the middle panel of Figure \ref{fig:all_obs}.
The bottom panel shows the data folded onto the fitted period, clearly demonstrating the sinusoidal nature of the signal.

\begin{table}
    \centering
    \begin{tabular}{lccc}
        Sector & Mid-MJD (d) & Period (d) & Amplitude (\%)  \\ \hline
        \textit{EUVE} & $\approx50000$ & $1.15\pm0.003$ & $\approx25$ \\
         15 & 58724   &$1.1459\pm0.0033$ & $0.127\pm0.016$ \\
         16 & 58750 &$1.1547\pm0.0051$& $0.102\pm0.018$ \\
        $15+16$ & 58736 & $1.1468\pm0.0014$ & $0.117\pm0.012$ \\
        Ephemeris (BJD):  & $2458737.560\pm0.018$ & \\
    \end{tabular}
    \caption{Measured periods and amplitudes for the variation at GD\,394. \textit{EUVE} measurements from \citet{dupuisetal00-1} are given as approximations as observations were obtained at multiple epochs with different instruments.} 
    \label{tab:peramp}
\end{table}

The periods and amplitudes for each sector are consistent to within 3$\sigma$, so we do not formally detect any amplitude or phase variability between the two sectors. However, the shape of the power spectrum is notably different between the two sectors. In particular, the amplitude of the $1.1547\pm0.0051$\,d signal in Sector 16 is weaker than the $1.1459\pm0.0033$ signal in Sector 15, and the $P/2$ harmonic is clearly detected in Sector 16 but not Sector 15.  

We combined the light curves from both sectors and repeated the analysis, finding a period of $1.1466\pm0.0015$\,days and an amplitude $0.117\pm0.012$\% in the {\em TESS} bandpass, centered at a wavelength of roughly 786.5\,nm. The ephemeris is defined as the peak of the model fit closest to the mid-point of the \textit{TESS} observations, $T_{\mathrm{Ephemeris}} =2458737.560\pm0.018$\,(BJD). Table \ref{tab:peramp} summarises the various period and amplitude measurements for GD\,394 from \textit{TESS} and \textit{EUVE}.



\begin{figure}
    \centering
    \includegraphics[width=\columnwidth]{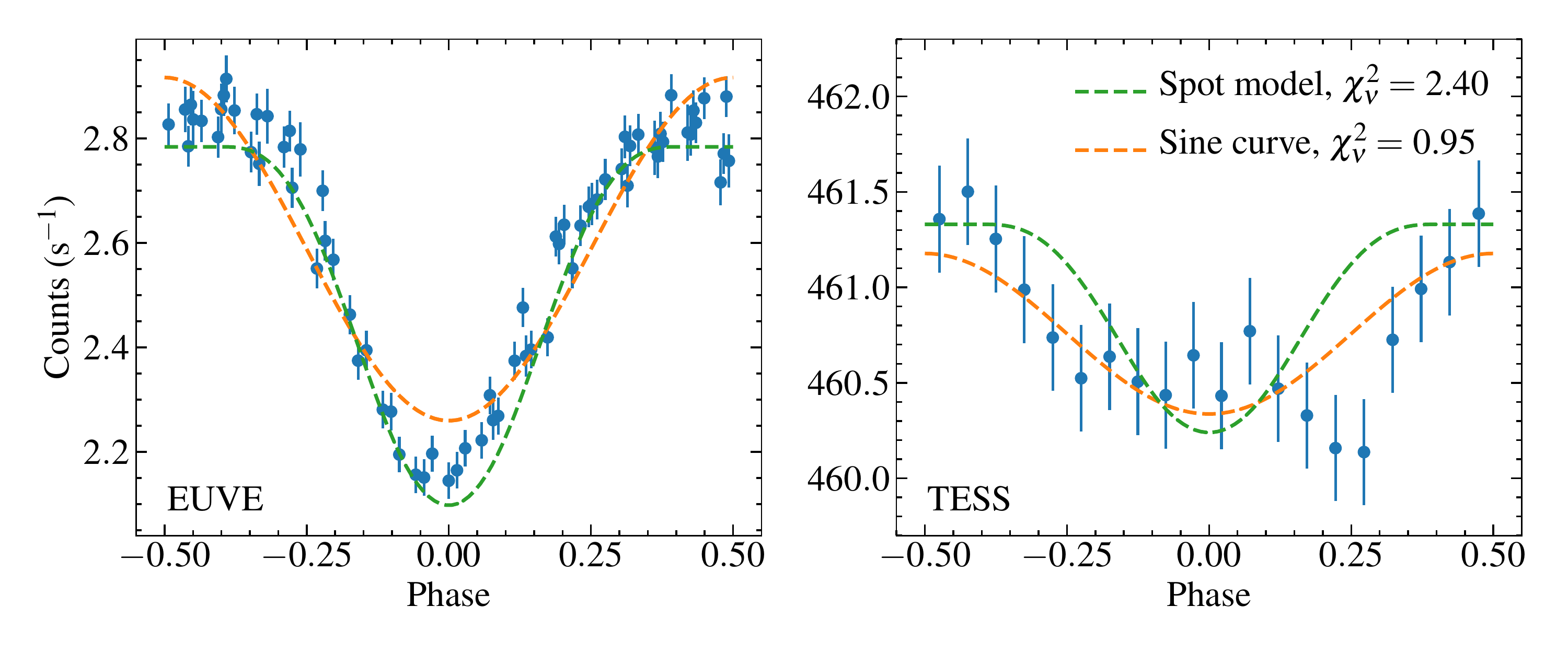}
    \caption{Comparison of the \textit{EUVE} and \textit{TESS} signals. Left: Recreation of Fig. 6 of \citet{dupuisetal00-1} showing the \textit{EUVE} DS light curve. Right: 2-sector \textit{TESS} light curve folded onto the measured 1.1468\,d period and binned to 20 points. In both panels the dashed green and orange lines show model fits to each light curve using the \citet{dupuisetal00-1} spot model and a sinusoidal respectively.}
    \label{fig:euve}
\end{figure}


\section{Discussion} \label{sec:disc}

The  $P = 1.145\pm 0.006$\,d periodic variation of GD\,394 in the {\em TESS} light curve is consistent with the $ 1.150\pm0.003$\,d period measured in the {\em EUVE} data by \citet{dupuisetal00-1}, with no strong evidence for period evolution in the roughly 24 years between the observations. On the other hand, the optical amplitude of $0.117\pm0.012$\% is much smaller than the $\approx$25\% variation in the EUV, and below the $\approx$1\,\% upper limits placed on FUV variation by \citet{wilsonetal19-1}.

Figure \ref{fig:euve} compares the folded \textit{EUVE} light curve and best-fit spot model from \citet{dupuisetal00-1} with the folded \textit{TESS} light curve. \citet{dupuisetal00-1} adopted a model where the spot is completely dark, with the ratio of the flux of the spot to the photosphere $kw =0$. To investigate whether the same spot is causing the \textit{TESS} variation, we fit a model keeping the geometry of the best-fit EUV model but varying the opacity. We find an $\approx$1\,\% flux ratio (i.e. $kw\approx0.99$ in the \citet{dupuisetal00-1} notation) provides a reasonable match to the folded \textit{TESS} light curveHowever, the sinusoidal model used to measure the period is statistically a better fit to the \textit{TESS} data despite being a poor model for the \textit{EUVE} light curve. The poor signal-to-noise ratio of the \textit{TESS} data dominates the fit, so confidently distinguishing between different models, and confirming whether the optical variation definitely has the same origin as the EUV variation, is not possible.






Optical variation was a prediction of the metal spot model favoured by \citet{dupuisetal00-1} as an explanation for the EUV variation. However, the metal spot model was not supported by the non-detection of changes in FUV metal absorption line strengths by \citet{wilsonetal19-1}. It is possible that the spot has a variable opacity, appearing strongly at the time of the 1993--1996 \textit{EUVE} observations, fading by the time of the 2015 \textit{HST} observations but returning in time to be observed with 2019 \textit{TESS} observations (with the period fixed by the white dwarf rotation period). However, this is likely too strong an appeal to coincidence, especially considering that Si absorption line strengths in the 2015 \textit{HST} spectra were consistent with the strengths of the same lines detected in \textit{HST} spectra obtained in 1992 by \citet{shipmanetal95-2}. Figure \ref{fig:euve} shows that a lower opacity version of the spot model fitted to the \textit{EUVE} data does does not exactly describe the \textit{TESS} light curve, raising the possibility that the geometry of the spot may have changed. If the \textit{TESS} mission is extended long enough for GD\,394 to be reobserved then tests for changes in the variation amplitude and shape will be possible. 

We are left requiring a mechanism that will generate flux variations of 25\% in the EUV, 0.117\% in the optical, and $\lesssim 1$\% in the FUV (the upper limit placed by light curves extracted from the time-tagged \textit{HST} spectroscopy by \citet{wilsonetal19-1}. The suggestion by \citet{veras+wolszczan19-1} that the variation is due to a hot spot at the base of a magnetic flux tube may fit these criteria, as a sufficiently hot spot could provide the required amplitude in the EUV, with the flux quickly dropping away in the Rayleigh–Jeans tail to the low levels observed at longer wavelengths. However, this would require spot temperatures of $\gtrsim 10^5$\,K, and thus far no  magnetic field has been detected in high-resolution spectroscopy of GD\,394 \citep[$B_e \leq 12$\,kG,][]{dupuisetal00-1, wilsonetal19-1}. The generation of a flux tube requires an orbiting metal-rich planetary fragment which may be radio-loud, and thus radio observations might provide an opportunity to test this model.
 

The various explanations for the flux variations at GD\,394, including metal spots, occultation by an outflow from an orbiting planet, or a magnetically induced hot spot,  could be tested by searching for phase differences between the two wavebands: Out of phase variation would favour a metal spot; in phase variation would point to a circumstellar cause or a hot spot. In practice, phasing up the {\em EUVE} and {\em TESS} observations is impossible given cycle-count ambiguities in the decades-long gap between them. Therefore, new, contemporaneous EUV and high-precision optical observations are required, although this will be challenging given the currently available observing facilities.

The \textit{TESS} detection strengthens the connection between GD\,394 and WD\,J1855+4207 suggested by \citet{hallakounetal18-1}, both stars having high ($> 30000$\,K) effective temperatures, high ionisation-level metal absorption lines with strengths well above those predicted by their effective temperatures, and weak, many-hour period optical modulation \citep{maozetal15-1}. Proposed future missions such as \textit{ESCAPE} \citep{franceetal19-1} could search for EUV variation at WD\,J1855+4207, confirming whether or not it is truly a GD\,394 analogue.


In conclusion, the \textit{TESS} observations of GD\,394 reveal the same 1.15\,d periodicity at optical wavelengths that was initially identified in the EUV more than two decades ago. The very small amplitude of the optical modulation, 0.12\,\%, explains why this signal remained undetected in previous ground-based observations of GD\,394. A physical explanation for the modulation now detected in two wavebands remains elusive.

\acknowledgments
We thank A. Vanderburg for useful advice regarding contamination from nearby stars in \textit{TESS}, and J. Dupuis for sharing the \textit{EUVE} light curves. JJH acknowledges support by the National Aeronautics and Space Administration through the \textit{TESS} Guest Investigator Program (80NSSC19K0378). This paper includes data collected by the \textit{TESS} mission. Funding for the \textit{TESS} mission is provided by the NASA Explorer Program.

%

\vspace{5mm}
\facilities{\textit{TESS} }

\defcitealias{astropy18-1}{Astropy Collaboration, 2018}

\software{Astropy \citepalias{astropy18-1}, Lightkurve \citepalias{lightkurve18-1}}





\bibliographystyle{aasjournal}
\bibliography{aabib}



\end{document}